\documentclass[10pt,doublesided, doublecolumn,final]{IEEEtran}
\usepackage{float}
\usepackage{algorithm}
\usepackage[noend]{algorithmic}
\usepackage[geometry]{ifsym}
\usepackage {ifsym}
\usepackage{amssymb}
\usepackage{yhmath}
\usepackage{amscd}
\usepackage{dsfont}
\usepackage{amsmath}
\usepackage{epsfig}
\usepackage{graphics}
\usepackage{psfrag}
\usepackage{rotating}
\usepackage{amsmath}
\usepackage{amsfonts}
\usepackage{url}
\usepackage{color}
\usepackage{float}
\usepackage{balance} 

\newtheorem{theorem}{Theorem}
\newtheorem{definition}{Definition}

\newcommand{\ind}[1]{\mathds{1}_{\left\lbrace #1 \right\rbrace}}

\newcommand{\bs}{\boldsymbol}

\newcommand{\ds}{\displaystyle}

\newcommand{\sfT}{\textsf{T}}

\newcommand{\Cldicnfb}{\mathcal{C} }

\begin{document}
\title{Noisy Channel-Output Feedback Capacity of the Linear Deterministic Interference Channel
%Decentralized Gaussian Interference Channels \\  with Degraded Output Feedback
%On the Impact of Noisy Feedback in Two-User Decentralized Interference Channels
}
\author{Victor Quintero, Samir M. Perlaza, Jean-Marie Gorce
\thanks{The authors are with the CITI Laboratory of the Institut National de Recherche en Informatique et en Automatique (INRIA), Universit\'{e} de Lyon and Institut National de Sciences Apliqu\'ees (INSA) de Lyon. 6 Av. des Arts 69621 Villeurbanne, France. ($\lbrace$Victor.Quintero-Florez, Samir. Perlaza, Jean-Marie.Gorce$\rbrace$@inria.fr).}
\thanks{Victor Quintero is also with Universidad del Cauca, Popay\'{a}n, Colombia.}
\thanks{Samir M. Perlaza is also with the Department of Electrical Engineering at Princeton University, Princeton, NJ.}
 \thanks{This research was supported in part by the European Commission under Marie Sklodowska-Curie Individual Fellowship No. 659316 (CYBERNETS); the Doctoral School in Electronics, Electrical Engineering and Automation (EEA) of Universit\'{e} de Lyon; Universidad del Cauca, Popay\'{a}n, Colombia; and the Administrative Department of Science, Technology and Innovation of Colombia (Colciencias), fellowship No. 617-2013.}
}

\maketitle

\begin{abstract}  
In this paper, the capacity region of the two-user linear deterministic (LD) interference channel with noisy output feedback (IC-NOF) is fully characterized.
This result allows the identification of several asymmetric scenarios in which implementing channel-output feedback in only one of the transmitter-receiver pairs is as beneficial as implementing it in both links, in terms of achievable individual rate and sum-rate improvements w.r.t. the case without feedback. In other scenarios, the use of channel-output feedback in any of the transmitter-receiver pairs benefits only one of the two pairs in terms of achievable individual rate improvements or simply, it turns out to be useless, i.e., the capacity regions with and without feedback turn out to be identical even in the full absence of noise in the feedback links.
%

%An achievability scheme, optimized for the IC-NOF, using standard tools such as message splitting, superposition coding and backward decoding is also presented.
\end{abstract}

  \section{Introduction}\label{SecIntroduction}
Perfect channel-output feedback (POF) has been shown to dramatically enlarge the capacity region of the two-user interference channel (IC) \cite{Tuninetti-ISIT-2007, Tuninetti-ITA-2010, Suh-TIT-2011, Vahid-TIT-2012,Yang-Tuninetti-TIT-2011}. More recently, the same observation has been made with a larger number of transmitter-receiver pairs in the IC \cite{Mohaher-TIT-2013}.
In general, when a transmitter observes the channel-output at its intended receiver, it obtains a noisy version of the sum of its own transmitted signal and the interfering signals from other transmitters. This implies that, subject to a finite feedback delay, transmitters know at least partially the information transmitted by other transmitters in the network. This induces an implicit cooperation between transmitters that allows them to use interference as side-information \cite{Tuninetti-ITA-2010, Yang-Tuninetti-TIT-2011, Prabhakaran-TIT-2011, Tuninetti-ITW-2012, Sahai-TIT-2013}. A more explicit cooperation is also observed in the case in which one of the transmitter-receiver pairs acts as a relay for the other transmitter-receiver pair by providing an alternative path: transmitter $i$ $\rightarrow$ receiver $j$ $\rightarrow$ transmitter $j$ $\rightarrow$ receiver $i$ \cite{Suh-TIT-2011}.  
These types of cooperation, even when it is not explicitly desired by both transmitter-receiver pairs,  play a fundamental role in enlarging the capacity region. Interestingly, this holds also in the case of fully decentralized networks in which each transmitter-receiver pair seeks exclusively to increase its individual rate. That is, channel-output feedback increases both the capacity region and the Nash equilibrium (NE) region \cite{Perlaza-TIT-2013}. 

Despite the vast existing literature, the benefits of feedback are unfortunately less well understood when the channel-output feedback links are impaired by additive noise. The capacity region  of the LD-IC with noisy channel-output feedback (NOF) is known only in the two-user symmetric case, see \cite{SyQuoc-TIT-2015}. The converse region in \cite{SyQuoc-TIT-2015} inherits existing outer bounds from the case of POF, the cut-set outer bounds and includes two new outer bounds. The outer-bounds inherited from the POF are those of the individual rates and the sum-rate in \cite{Suh-TIT-2011}. The new outer-bounds are of the form $R_1 + R_2$ and $R_i + 2 R_j$. 
The achievability region in \cite{SyQuoc-TIT-2015} is obtained using a particularization of the achievability scheme presented in \cite{Tuninetti-ISIT-2007}, which holds for a more general model, i.e., interference channel with generalized feedback. 

In this paper, the results presented in \cite{SyQuoc-TIT-2015} are generalized for the asymmetric case and the corresponding capacity region of the two-user LD-IC-NOF is fully characterized. This generalization is achieved by using the same tools as in \cite{SyQuoc-TIT-2015}, however, it is far from trivial due to the number of parameters that describe this channel model: two forward signal to noise ratios (SNRs) $\overrightarrow{n}_{11}, \overrightarrow{n}_{22}$, two feedback SNRs $\overleftarrow{n}_{11} , \overleftarrow{n}_{22} $ and two forward interference to noise ratios (INRs) $n_{12}, n_{21}$.   

The new converse region also inherits existing outer bounds from the case of POF, the cut-set outer bounds and includes two new outer bounds on $R_1 + R_2$ and $R_i + 2 R_j$. These new bounds generalize those presented in \cite{SyQuoc-TIT-2015}.
The achievability region is obtained by using a coding scheme that combines a three-part message splitting, superposition coding and backward decoding. Despite the fact that this coding scheme is built using the exact number of required message-splitting parts for the IC-NOF, it can still be considered  as a special case of the general scheme presented in \cite{Tuninetti-ISIT-2007}. 

Finally, this paper is concluded by a discussion in which numerical examples are presented to highlight the benefits of NOF. At the same time, examples in which NOF is absolutely useless in terms of capacity region improvement are also presented.
Due to space constraints, the achievability and the converse parts of the proof are presented in \cite{Quintero-INRIA-TechRep-2015}.%
\section{Linear Deterministic Interference Channel with Noisy-Channel Output Feedback}
\begin{figure}[t!]
 \centerline{\epsfig{figure=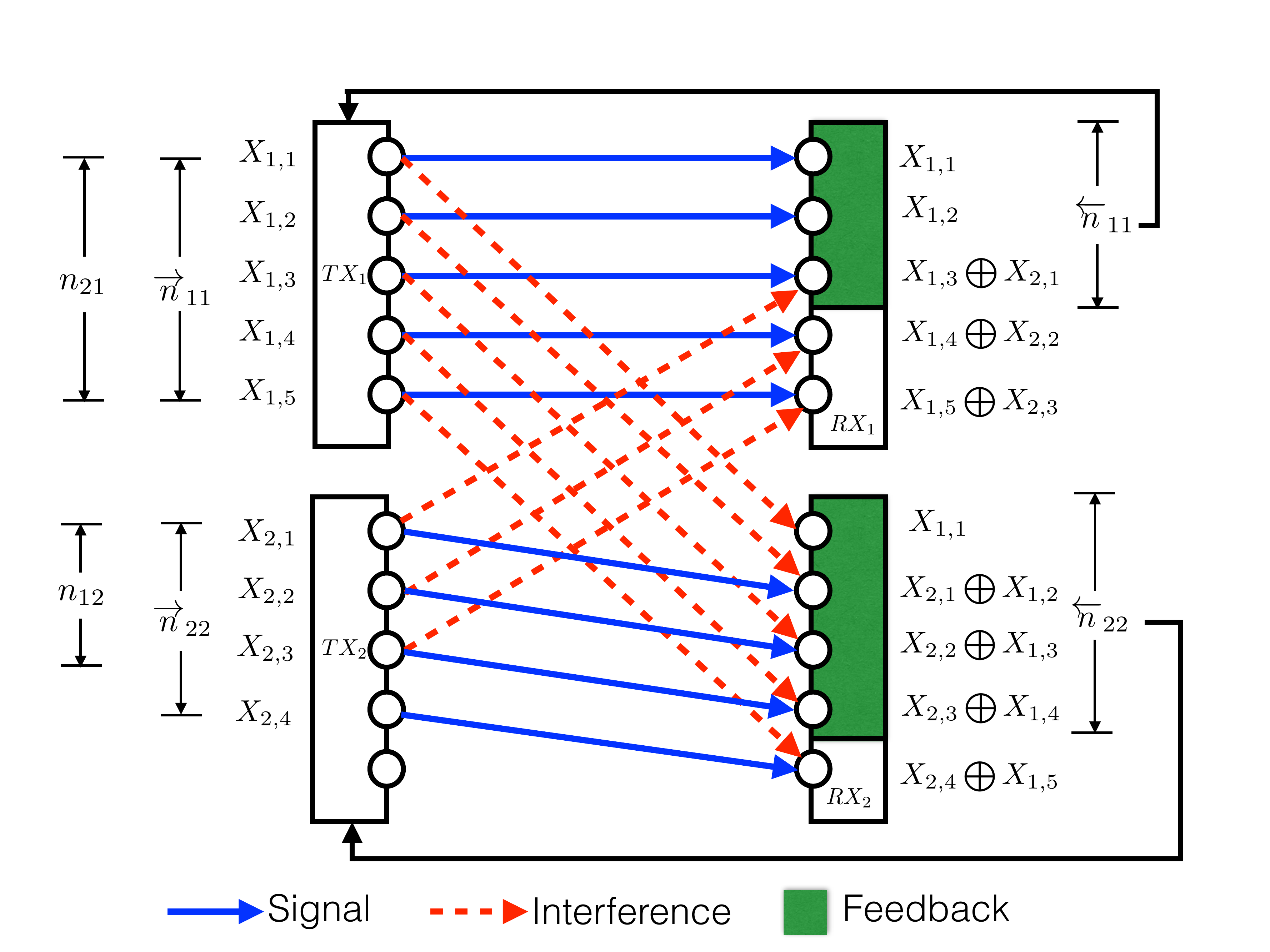,width=.42\textwidth}}
% \vspace{-0.3in}
  \caption{Two-user linear deterministic interference channel with noisy channel-output feedback (LD-IC-NOF).}
  \label{FigLD-IC-NOF}
\end{figure}

Consider the two-user LD-IC-NOF, with parameters $\overrightarrow{n}_{11}$, $\overrightarrow{n}_{22}$, $n_{12}$, $n_{21}$, $\overleftarrow{n}_{11}$ and $\overleftarrow{n}_{22}$ described in Fig.~\ref{FigLD-IC-NOF}. 
$\overrightarrow{n}_{ii}$, $i \in \lbrace1,2\rbrace$, is a non-negative integer used to represent the signal to noise ratio (SNR) at receiver $i$; $n_{ij}$, $i \in \lbrace1,2\rbrace$ and $j \in \lbrace1,2 \rbrace \setminus\lbrace i \rbrace$, is a non-negative integer used to represent the interference to noise ratio (INR) at receiver $i$ from transmitter $j$; and $\overleftarrow{n}_{ii}$, $i \in \lbrace1,2\rbrace$, is a non-negative integer used to represent the signal to noise ratio (SNR) at transmittter $i$ in the feedback link from receiver $i$. 
At transmitter $i$, with $i \in \lbrace 1, 2 \rbrace$, the channel-input $\bs{X}_{i}^{(n)}$ at channel use $n$, with $n \in \lbrace 1, \ldots, N \rbrace$, is a $q$-dimensional binary vector $\bs{X}_{i}^{(n)} = \left(X_{i,1}^{(n)}, \ldots, X_{i,q}^{(n)}\right)^{\sfT}$, with  $q=\ds\max \left(\overrightarrow{n}_{11}, \overrightarrow{n}_{22}, n_{12}, n_{21}\right)$  and $N$ the block-length. 
At receiver $i$, the channel-output $\overrightarrow{\bs{Y}}_{i}^{(n)}$ at channel use $n$ is also a $q$-dimensional binary vector $\overrightarrow{\bs{Y}}_{i}^{(n)} = \left(\overrightarrow{Y}_{i,1}^{(n)}, \ldots, \overrightarrow{Y}_{i,q}^{(n)}\right)^{\sfT}$. 
The input-output relation during channel use $n$ is given as follows
\begin{IEEEeqnarray}{lcl}
\label{EqLDICsignals}
\overrightarrow{\bs{Y}}_{i}^{(n)}&=& \bs{S}^{q - \overrightarrow{n}_{ii}} \bs{X}_{i}^{(n)} + \bs{S}^{q - n_{ij}} \bs{X}_{j}^{(n)},
\end{IEEEeqnarray}
and the feedback signal  available at transmitter $i$ at the end of channel use $n$ is:
\begin{IEEEeqnarray}{lcl}\label{EqLDICsignalsc}
\bs{\overleftarrow{Y}}_{i}^{(n)} &=& \bs{S}^{(q-\overleftarrow{n}_{ii})^+} \, \bs{\overrightarrow{Y}}_{i}^{(n-d)},
\end{IEEEeqnarray}
where $d$ is a finite feedback delay, additions and multiplications are defined over the binary field, and $\bs{S}$ is a $q\times q$ lower shift matrix. % of the form
%\begin{align}
%\bs{S}=
%\left[
%  \begin{array}{cccccc}
%    0 & 0 &0 & \cdots & 0 \\
%    1 & 0 & 0 &\cdots & 0\\
%    0 & 1 & 0 &\cdots  & \vdots\\
%    \vdots & \ddots & \ddots  & \ddots & 0\\
%    0 & \cdots & 0 & 1& 0
%  \end{array}
%\right].\nonumber
%\end{align}

The parameters $\overrightarrow{n}_{ii}$, $\overleftarrow{n}_{ii}$ and $n_{ij}$ correspond to $\left \lfloor\frac{1}{2}\log_2\left(\overrightarrow{\textrm{SNR}}_i\right) \right \rfloor$, $\left \lfloor \frac{1}{2} \log_2 \left(\overleftarrow{\textrm{SNR}}_i\right) \right \rfloor$ and  $\left \lfloor \frac{1}{2} \log_2 \left(\textrm{INR}_{ij}\right) \right \rfloor$ respectively, where $\overrightarrow{\textrm{SNR}}_i$, $\overleftarrow{\textrm{SNR}}_i$ and $\textrm{INR}_{ij}$ are parameters of the Gaussian interference channel (G-IC).

Transmitter $i$ sends $M_i$ information bits $b_{i,1}, \ldots, b_{i,M_i}$ by sending the codeword $\left(\bs{X}_{i}^{(1)}, \ldots, \bs{X}_{i}^{(N)}\right)$. 
The encoder of transmitter $i$ can be modeled as a set of deterministic mappings $f_i^{(1)}, \ldots, f_i^{(N)}$, with $f_i^{(1)}: \lbrace 0,1 \rbrace^{M_i}  \rightarrow \lbrace 0, 1 \rbrace^{q}$ and $\forall n \in \lbrace 2, \ldots, N\rbrace$, $f_i^{(n)}: \lbrace 0,1 \rbrace^{M_i}  \times \lbrace 0, 1 \rbrace^{q(n-1)} \rightarrow \lbrace 0, 1 \rbrace^{q}$, such that 
\begin{IEEEeqnarray}{lcl}\label{EqEnconderf}
\bs{X}_{i}^{(1)} &=& f_i^{(1)}\big(b_{i,1},\ldots,b_{i,M_i} \big) \mbox{ and  }\\
\bs{X}_{i}^{(n)} &=& f_i^{(n)}\big(b_{i,1},\ldots,b_{i,M_i}, \bs{\overleftarrow{Y}}_{i}^{(1)}, \ldots, \bs{\overleftarrow{Y}}_{i}^{(n-1)} \big). 
\end{IEEEeqnarray}
At the end of the block, receiver $i$ uses the sequence $\bs{Y}_{i}^{(1)}, \ldots, \bs{Y}_{i}^{(N)}$ to  generate the estimates $\hat{b}_{i,1}, \ldots, \hat{b}_{i,M_i}$.  
The average bit error probability at receiver $i$, denoted by $p_i$, is calculated as follows
\begin{equation}
\label{EqPe}
p_i = \frac{1}{M_i} \ds\sum_{\ell = 1}^{M_i} \ind{\hat{b}_{i,\ell} \neq {b}_{i,\ell} }.
\end{equation}
A rate pair $(R_1, R_2) \in \mathds{R}_{+}^{2}$ is said to be achievable if it satisfies the following definition.
\begin{definition}[Achievable Rate Pairs]\label{DefAchievableRatePairs}\empty{
The rate pair $(R_1, R_2) \in \mathds{R}_{+}^{2}$ is achievable if there exists at least one pair of codebooks $\mathcal{X}_1$ and $\mathcal{X}_2$ with codewords of length $N$,  with the corresponding encoding functions $f_1^{(1)}, \ldots, f_1^{(N)}$ and $f_2^{(1)}, \ldots, f_2^{(N)}$ such that the average bit error probability can be made arbitrarily small by letting the block length $N$ grows to infinity.
}
\end{definition}

The following section determines the set of all the rate pairs $(R_1,R_2)$ that are achievable in the LD-IC-NOF with parameters $\overrightarrow{n}_{11}$, $\overrightarrow{n}_{22}$, $n_{12}$, $n_{21}$, $\overleftarrow{n}_{11}$ and $\overleftarrow{n}_{22}$.

\section{Main Results}
Denote by $\Cldicnfb(\overrightarrow{n}_{11}, \overrightarrow{n}_{22}, n_{12}, n_{21}, \overleftarrow{n}_{11} , \overleftarrow{n}_{22})$ the capacity region of the LD-IC-NOF with parameters $\overrightarrow{n}_{11}$, $\overrightarrow{n}_{22}$, $n_{12}$, $n_{21}$, $\overleftarrow{n}_{11}$ and $\overleftarrow{n}_{22}$. Theorem~\ref{TheoremANFBLDMCap} (on top of next page) fully characterizes the capacity region $\Cldicnfb(\overrightarrow{n}_{11}, \overrightarrow{n}_{22}, n_{12}, n_{21}, \overleftarrow{n}_{11} , \overleftarrow{n}_{22})$.
\begin{figure*}[t]
\begin{theorem}\label{TheoremANFBLDMCap} \emph{
The capacity region $\Cldicnfb(\overrightarrow{n}_{11}, \overrightarrow{n}_{22}, n_{12}, n_{21}, \overleftarrow{n}_{11} , \overleftarrow{n}_{22})$ of the two-user LD-IC-NOF is the set of non-negative rate pairs $(R_1,R_2)$ that satisfy $ \forall i \in \lbrace 1, 2 \rbrace$ and $j\in\lbrace 1, 2 \rbrace\setminus\lbrace i \rbrace$:
\begin{IEEEeqnarray}{ll}
\label{EqRegionCwnFBa}
R_{i}& \leqslant \min\left(\max\left(\overrightarrow{n}_{ii},n_{ji}\right),\max\left(\overrightarrow{n}_{ii},n_{ij}\right)\right), \\
\label{EqRi-2}
R_i& \leqslant \min\left(\max\left(\overrightarrow{n}_{ii},n_{ji}\right),\max\left(\overrightarrow{n}_{ii},\overleftarrow{n}_{jj}-\left(\overrightarrow{n}_{jj}-n_{ji}\right)^+\right)\right),\\
\label{EqRi+Rj-1}
R_1&+R_2  \leqslant \min \left( \max \left(\overrightarrow{n}_{11},n_{12}\right)+\left(\overrightarrow{n}_{22}-n_{12}\right)^+, \max \left(\overrightarrow{n}_{22},n_{21}\right)+\left(\overrightarrow{n}_{11}-n_{21}\right)^+  \right), \\
\label{EqRi+Rj-2}
R_1&+R_2 \leqslant \max\left(\left(\overrightarrow{n}_{11}-n_{12}\right)^+,n_{21}\right)+\max\left(\left(\overrightarrow{n}_{22}-n_{21}\right)^+,n_{12}\right)\\
\nonumber
 &+\Big(\left(\min\left(\overleftarrow{n}_{11},\max\left(\overrightarrow{n}_{11},n_{12}\right)\right)-\left(\overrightarrow{n}_{11}-{n}_{12} \right)^+\right)^+-\left(n_{12}-\overrightarrow{n}_{11}\right)^+-\min\left(\overrightarrow{n}_{11},{n}_{21}\right)+\min\left(\left(\overrightarrow{n}_{11}-n_{12}\right)^+,n_{21}\right)\Big)^+\\
 \nonumber
& +\Big(\left(\min\left(\overleftarrow{n}_{22},\max\left(\overrightarrow{n}_{22},n_{21}\right)\right)-\left(\overrightarrow{n}_{22}-{n}_{21} \right)^+\right)^+-\left(n_{21}-\overrightarrow{n}_{22}\right)^+-\min\left(\overrightarrow{n}_{22},{n}_{12}\right)+\min\left(\left(\overrightarrow{n}_{22}-n_{21}\right)^+,n_{12}\right)\Big)^+,\\
\label{Eq2Ri+Rj}
2R_i&+R_j  \leqslant \max\left(\overrightarrow{n}_{jj},n_{ji}\right)+\max\left(\overrightarrow{n}_{ii},{n}_{ij} \right)+\left(\overrightarrow{n}_{ii}-{n}_{ji} \right)^+-\min\left(\left(\overrightarrow{n}_{jj}-n_{ji}\right)^+,n_{ij}\right)\\
\nonumber
 & +\Big(\left(\min\left(\overleftarrow{n}_{jj},\max\left(\overrightarrow{n}_{jj},n_{ji}\right)\right)-\left(\overrightarrow{n}_{jj}-{n}_{ji} \right)^+\right)^+-\left(n_{ji}-\overrightarrow{n}_{jj}\right)^+-\min\left(\overrightarrow{n}_{jj},{n}_{ij}\right)+\min\left(\left(\overrightarrow{n}_{jj}-n_{ji}\right)^+,n_{ij}\right)\Big)^+,\\
\end{IEEEeqnarray}
}
\end{theorem} 
\end{figure*}

\subsection{Proofs}
Theorem~\ref{TheoremANFBLDMCap} fully characterizes the capacity region of the LD-IC-NOF. That is, the converse and achievable regions are identical. 
In the converse region, the inequalities \eqref{EqRegionCwnFBa} and \eqref{EqRi+Rj-1} are inherited from the converse region of the LD-IC-POF in \cite{Suh-TIT-2011}. The inequality in \eqref{EqRi-2} is a simple cut-set bound whose proof is presented in \cite{Quintero-INRIA-TechRep-2015}.
The inequalities \eqref{EqRi+Rj-2} and \eqref{Eq2Ri+Rj} are new and generalize those presented in \cite{SyQuoc-TIT-2015}. The corresponding proofs are presented in \cite{Quintero-INRIA-TechRep-2015}.

The achievability region is obtained using a coding scheme that combines a three-part message splitting, superposition coding and backward decoding, as first suggested in \cite{Tuninetti-ISIT-2007, Suh-TIT-2011, Yang-Tuninetti-TIT-2011}. This coding scheme is fully described in \cite{Quintero-INRIA-TechRep-2015} and it is specially designed for the IC-NOF.  However, it can also be obtained as a special case of the more general scheme, i.e., interference channel with generalized feedback, presented in \cite{Tuninetti-ISIT-2007}. The relevance of this new achievability scheme is that it plays a key role in the achievability of the NE region, subject to the inclusion of random messages, as suggested in \cite{Perlaza-TIT-2013, Berry-TIT-2011}. Nonetheless, the analysis of the achievability of the NE region \cite{Perlaza-ISIT-2014a} is out of the scope of this paper.

\subsection{Connections to Existing Results}

In the case in which channel-output feedback is not available, i.e., $\overleftarrow{n}_{11} = \overleftarrow{n}_{22}=0$, $\overrightarrow{n}_{ii} = n_{ii}$, for all $i \in \lbrace 1 ,2 \rbrace$,  Theorem~\ref{TheoremANFBLDMCap} reduces to the capacity region of the LD-IC without feedback, i.e.,  Lemma $4$ in \cite{Bresler-ETT-2008}. 

Conversely, when perfect channel-output feedback links are available i.e., ${\overleftarrow{n}_{11}=\max\left(\overrightarrow{n}_{11},{n}_{12}\right)}$ and ${\overleftarrow{n}_{22}=\max\left(\overrightarrow{n}_{22}, {n}_{21}\right)}$,  Theorem~\ref{TheoremANFBLDMCap} reduces to the capacity of the LD-IC-POF, that is, Corollary~$1$ in \cite{Suh-TIT-2011} in the corresponding notation. 

In the symmetric case in which channel-output feedback links are impaired by noise, i.e., ${\overrightarrow{n}_{11}=\overrightarrow{n}_{22}=n}$, ${n_{12}=n_{21}=m}$ and ${\overleftarrow{n}_{11}=\overleftarrow{n}_{22}=\ell}$,  Theorem~\ref{TheoremANFBLDMCap} reduces to the capacity of the symmetric  LD-IC-NOF, i.e., Theorem~$1$ in \cite{SyQuoc-TIT-2015}.

%It is also worth noting that the outer-bounds on the sum-rate in   Theorem~\ref{TheoremANFBLDMCap} for the case in which there exist source-cooperation, e.g., $\overrightarrow{n}_{11}=n_{1,3}$, $\overrightarrow{n}_{22}=n_{2,4}$, $n_{12}=n_{2,3}$, $n_{21}=n_{1,4}$ and $\overleftarrow{n}_{11}=\overleftarrow{n}_{22}=n_{1,2}=n_{2,1}=n_{c}$, reduce to the sum-rate outer bound presented in Theorem $1$ in \cite{Prabhakaran-TIT-2011}.

Note also that in the case in which the interference channel is symmetric, Theorem~\ref{TheoremANFBLDMCap} partially generalizes Theorem $4.1$ in \cite{Sahai-TIT-2013}. For instance, note that when there exists only one perfect channel-output feedback, i.e., ${\overrightarrow{n}_{11}=\overrightarrow{n}_{22}=n}$, ${n_{12}=n_{21}=m}$,  ${\overleftarrow{n}_{11}=\max \left(\overrightarrow{n}_{11},n_{12}\right)}$, and ${\overleftarrow{n}_{22}=0}$,  Theorem~\ref{TheoremANFBLDMCap} reduces to Theorem $4.1$ (model $1000$) in \cite{Sahai-TIT-2013}.

In the two-user IC-NOF, a transmitter sees a noisy version of the sum of its own transmitted signal and the interfering signal from the other transmitter. Hence, subject to a finite delay, one transmitter knows at least partially the information transmitted by the other transmitter in the network. This observation highlights the connections between the IC with feedback and the IC with source cooperation studied in \cite{Prabhakaran-TIT-2011}.   

\subsection{Discussion}

This section provides a set of numerical examples in which particular scenarios are highlighted to show that channel-output feedback can be strongly beneficial for enlarging the capacity region of the two-user LD-IC even when channel-output feedback links are impaired by noise. At the same time, it also highlights other examples in which channel-output feedback does not bring any benefit in terms of the capacity region. These benefits are given in terms of the following metrics: $(a)$ individual rate improvements $\Delta_1$ and $\Delta_2$; and $(b)$ sum-rate improvement $\Sigma$. 

In order to formally define $\Delta_1$, $\Delta_2$ and $\Sigma$, consider an LD-IC-NOF with parameters $\overrightarrow{n}_{11}$, $\overrightarrow{n}_{22}$, $n_{12}$, $n_{21},$ $\overleftarrow{n}_{11}$ and $\overleftarrow{n}_{22}$.
%
 %\footnote{This is an abbreviation of $\Cldicnfb(\overrightarrow{n}_{11}, \overrightarrow{n}_{22}, n_{12}, n_{21},\overleftarrow{n}_{11} , \overleftarrow{n}_{22})$.}. 
%
The maximum improvement $\Delta_i(\overrightarrow{n}_{11}, \overrightarrow{n}_{22}, n_{12}, n_{21},\overleftarrow{n}_{11} , \overleftarrow{n}_{22})$ of the individual rate $R_i$ due to the effect of channel-output feedback with respect to the case without feedback is:
\begin{IEEEeqnarray}{lcl}
\Delta_{i}(\overrightarrow{n}_{11}, \overrightarrow{n}_{22}, n_{12}, &n_{21} &, \overleftarrow{n}_{11} , \overleftarrow{n}_{22} )  = \\
\nonumber
& \max_{R_{j} > 0} & \sup_{\begin{array}{l} 	(R_i, R_j) \in \Cldicnfb_{1} \\ 
								(R_i^{\dagger}, R_j)  \in \Cldicnfb_{2} \\  \end{array} } R_{i} - R_i^{\dagger},
\end{IEEEeqnarray}
and the maximum sum rate improvement $\Sigma(\overrightarrow{n}_{11}, \overrightarrow{n}_{22}, n_{12}, n_{21},\overleftarrow{n}_{11} , \overleftarrow{n}_{22})$ with respect to the case without feedback is:
\begin{IEEEeqnarray}{lcl}
\Sigma(\overrightarrow{n}_{11}, \overrightarrow{n}_{22}, & n_{12} &,  n_{21}, \overleftarrow{n}_{11}, \overleftarrow{n}_{22} )  = \\
\nonumber
& & \sup_{\begin{array}{l} 	(R_1, R_2) \in \Cldicnfb_{1} \\ 
								(R_1^{\dagger}, R_2^{\dagger})  \in \Cldicnfb_{2} \\  \end{array} } R_{1}+R_{2} - (R_1^{\dagger}+R_2^{\dagger}),
\end{IEEEeqnarray}
where ${\Cldicnfb_{1} = \Cldicnfb(\overrightarrow{n}_{11}, \overrightarrow{n}_{22}, n_{12}, n_{21},\overleftarrow{n}_{11} , \overleftarrow{n}_{22})} $ and ${\Cldicnfb_{2} = \Cldicnfb(\overrightarrow{n}_{11}, \overrightarrow{n}_{22}, n_{12}, n_{21},0,0)}$ are the capacity region with noisy channel-output feedback and without feedback, respectively. The following describes particular scenarios that highlight some interesting observations.

\subsubsection{Example 1: Only one channel-output feedback link allows simultaneous maximum improvement of both individual rates}\label{SecExample1}
\begin{figure}[t!]
\centerline{\epsfig{figure=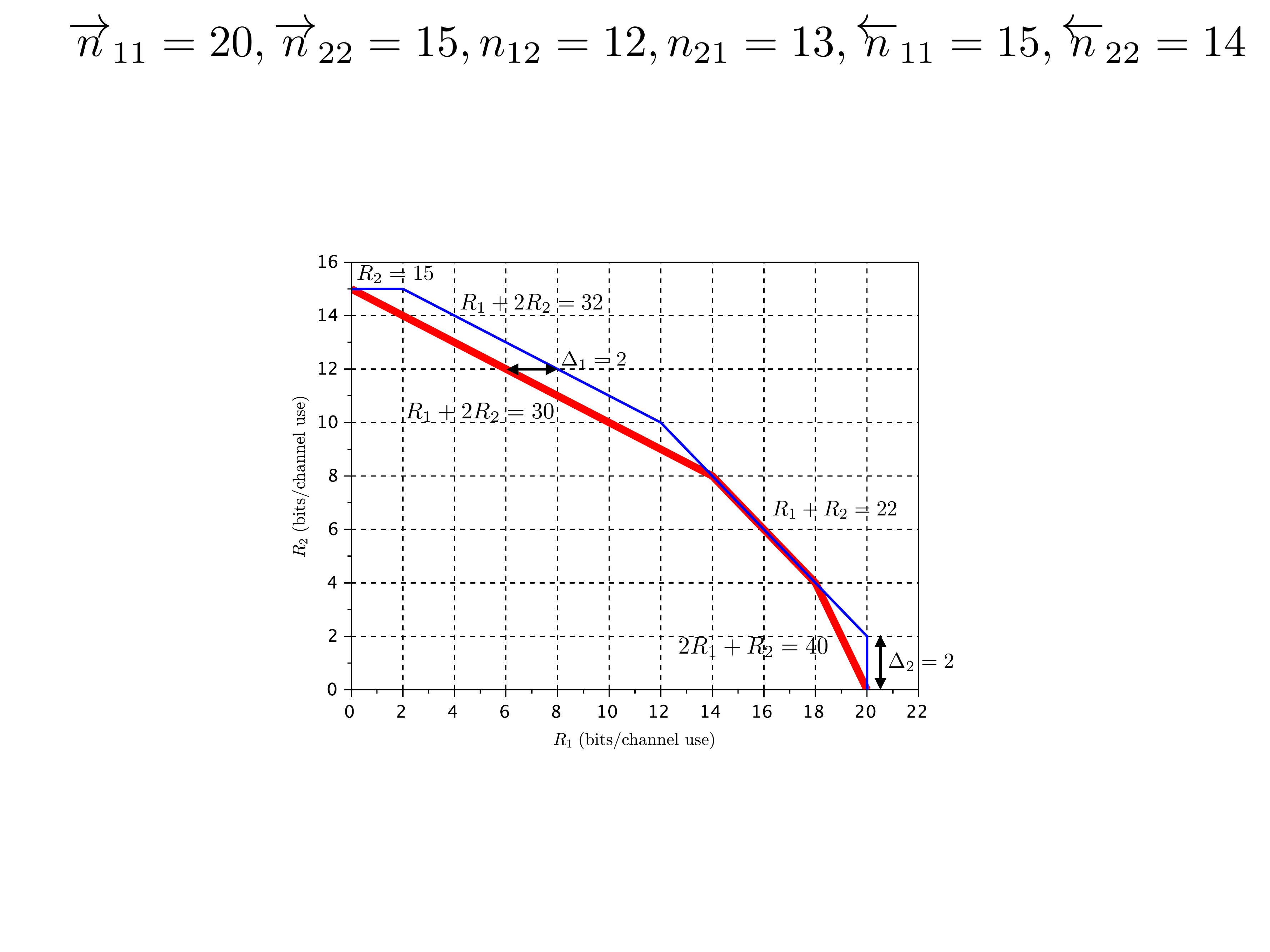,width=0.34\textwidth}}
\caption{Capacity region $\Cldicnfb(20,15,12,13,0,0)$ without feedback  (thick red line) and $\Cldicnfb(20,15,12,13,15,14)$ with noisy channel-output feedback (thin blue line) of the example in Sec.~\ref{SecExample1}. Note that $\Delta_{1}(20,15,12,13,15,14) = 2$ bits/ch.use, $\Delta_{2}(20,15,12,13,15,14) = 2$ bits/ch.use and $\Sigma(20,15,12,13,15,14) = 0$ bits/ch.use.} 
\label{FigExample1capb}
\end{figure}
\begin{figure}[t!]
\centerline{\epsfig{figure=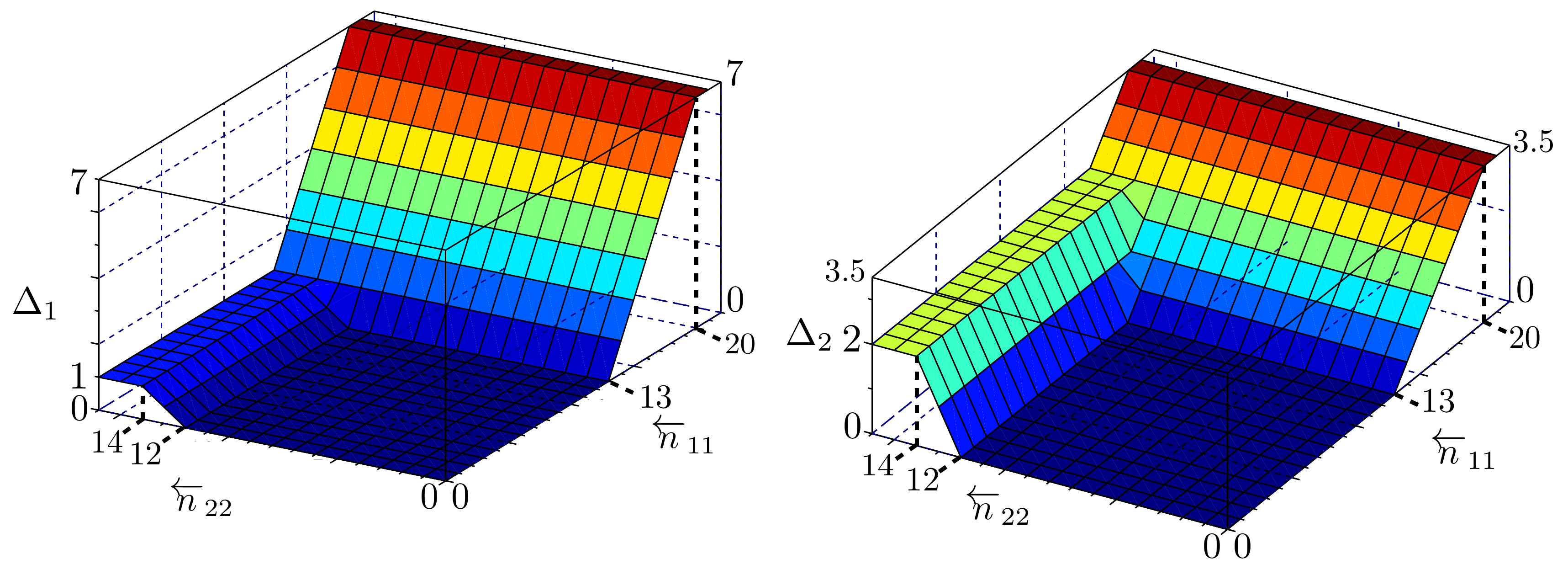,width=.43\textwidth}}
\caption{Maximum improvement of individual rates of the example in Sec.~\ref{SecExample1}
%$\Delta_{i}(\overrightarrow{n}_{11}, \overrightarrow{n}_{22}, n_{12}, n_{21}, \overleftarrow{n}_{11} ,\overleftarrow{n}_{22})$, with $i \in \lbrace 1,2\rbrace$, as a function of $\overleftarrow{n}_{11}$ and $\overleftarrow{n}_{22}$ and fixed parameters $\overrightarrow{n}_{11} = $, $\overrightarrow{n}_{22} = $, $n_{12} = $, $n_{21} = $.
}
\label{FigExample1a}
\end{figure}
Consider the case in which transmitter-receiver pairs $1$ and $2$ are in weak and moderate interference regimes, with $\overrightarrow{n}_{11} = 20$, $\overrightarrow{n}_{22} = 15$, $n_{12} = 12$, $n_{21} = 13$. 
In Fig.~\ref{FigExample1capb} the capacity region is plotted without channel-output feedback and with noisy channel-output feedback (${\overleftarrow{n}_{11}=15}, {\overleftarrow{n}_{22}=14}$).
In Fig.~\ref{FigExample1a}, $\Delta_{i}(20, 15, 12, 13, \overleftarrow{n}_{11} ,\overleftarrow{n}_{22})$ is plotted for both $i = 1$ and $i=2$ as a function of $\overleftarrow{n}_{11}$ and  $\overleftarrow{n}_{22}$. Therein, it is shown that:
$(a)$ Increasing parameter $\overleftarrow{n}_{11}$ beyond threshold $\overleftarrow{n}_{11}^*=13$ allows simultaneous improvement of both individual rates independently of the value of $\overleftarrow{n}_{22}$. Note that in the case of perfect channel-output feedback, i.e., ${\overleftarrow{n}_{11}= \max\left(\overrightarrow{n}_{11} , n_{12} \right)}$, the maximum improvement of both individual rates is simultaneously achieved even when ${\overleftarrow{n}_{22}=0}$.
$(b)$ Increasing parameter $\overleftarrow{n}_{22}$ beyond threshold $\overleftarrow{n}_{22}^*=12$  provides simultaneous improvement of both individual rates. However, the improvement on the individual rate $R_2$ strongly depends on the value of $\overleftarrow{n}_{11}$.
$(c)$ Finally, the sum rate does not increase by using channel-output feedback in this case.

\subsubsection{Example 2: Only one channel-output feedback link allows maximum improvement of one individual rate and the sum-rate}\label{SecExample2}
\begin{figure}[t!]
\centerline{\epsfig{figure=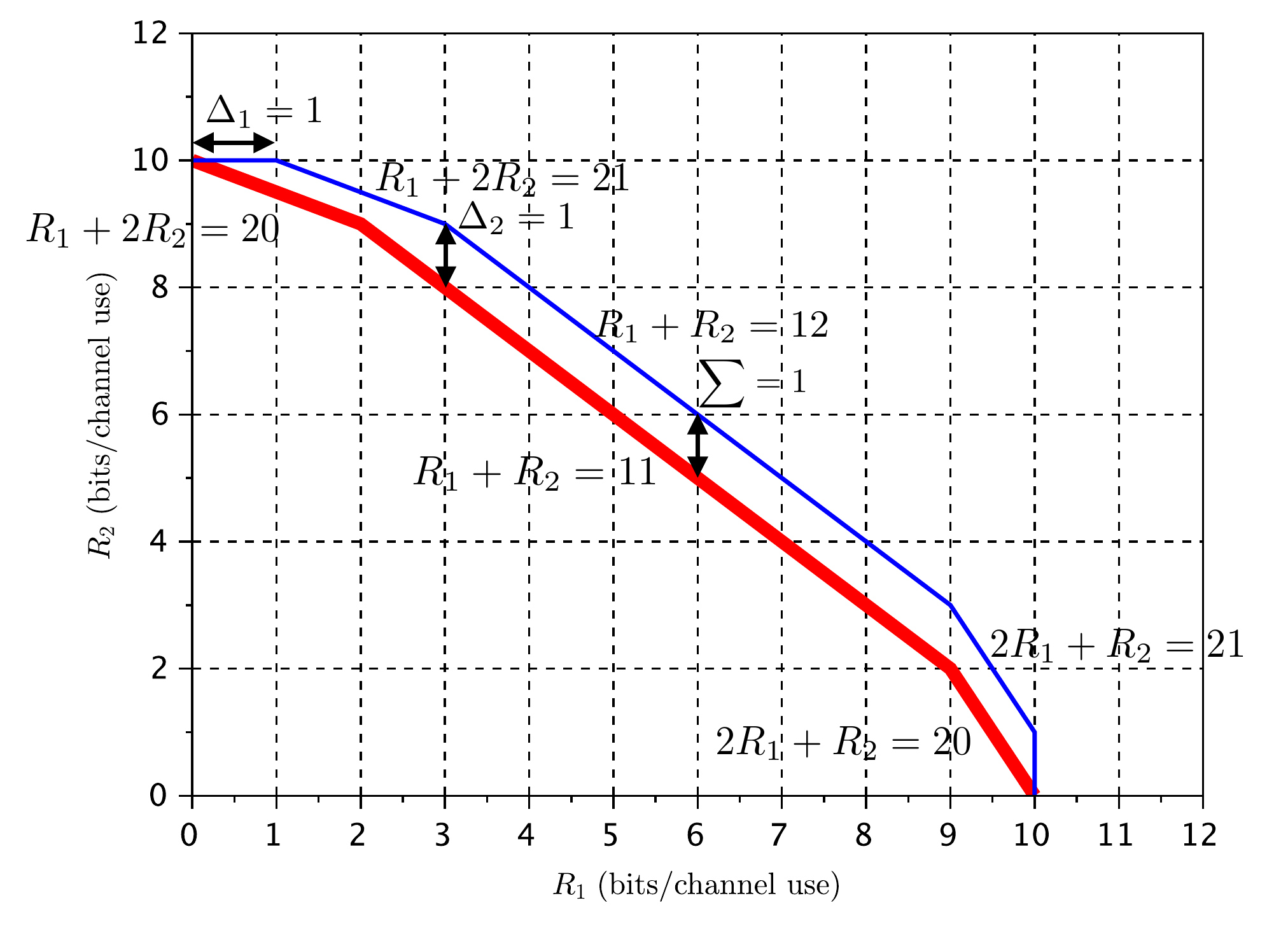,width=0.35\textwidth}}
\caption{Capacity region $\Cldicnfb(10,10,3,8,0,0)$ without feedback (thick red line) and  $\Cldicnfb(10,10,3,8,9,4)$ with noisy channel-output feedback (thin blue line) of the example in Sec.~\ref{SecExample2}. Note that $\Delta_{1}(10,10,3,8,9,4) = 1$ bit/ch.use, $\Delta_{2}(10,10,3,8,9,4) = 1$ bit/ch.use and $\Sigma(10,10,3,8,9,4) = 1$ bit/ch.use.}
\label{FigExample2capb}
\end{figure}
\begin{figure}[h]
\centerline{\epsfig{figure=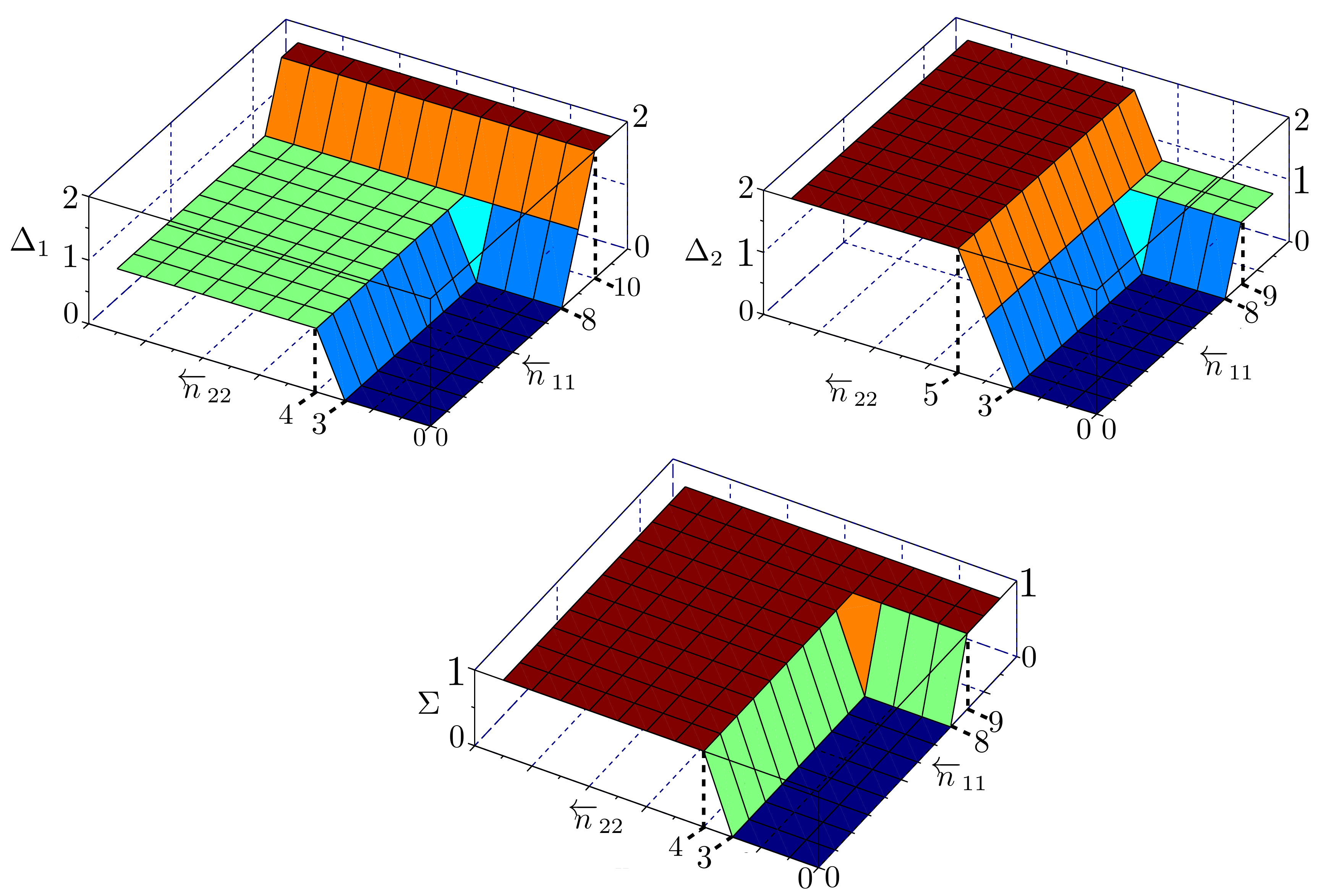,width=.43\textwidth}}
\caption{Maximum improvement of one individual rate and the sum rate of the example in Sec.~\ref{SecExample2}
%$\Delta_{i}(\overrightarrow{n}_{11}, \overrightarrow{n}_{22}, n_{12}, n_{21}, \overleftarrow{n}_{11} ,\overleftarrow{n}_{22})$, with $i \in \lbrace 1,2\rbrace$, as a function of $\overleftarrow{n}_{11}$ and $\overleftarrow{n}_{22}$ and fixed parameters $\overrightarrow{n}_{11} = $, $\overrightarrow{n}_{22} = $, $n_{12} = $, $n_{21} = $.
}
\label{FigExample2a}
\end{figure}
Consider the case in which transmitter-receiver pairs $1$ and $2$ are in very weak and moderate interference regimes, with $\overrightarrow{n}_{11} = 10$, $\overrightarrow{n}_{22} = 10$, $n_{12} = 3$, $n_{21} = 8$. 
In Fig.~\ref{FigExample2capb} the capacity region is plotted without channel-output feedback and with noisy channel-output feedback($\overleftarrow{n}_{11}=9, \overleftarrow{n}_{22}=4$).
In Fig.~\ref{FigExample2a}, $\Delta_{i}(10, 10, 3, 8, \overleftarrow{n}_{11} ,\overleftarrow{n}_{22})$ is plotted for both $i = 1$ and $i=2$ as a function of $\overleftarrow{n}_{11}$ and  $\overleftarrow{n}_{22}$. 
Therein, it is shown that: 
$(a)$  Increasing $\overleftarrow{n}_{11}$ beyond threshold $\overleftarrow{n}_{11}^*=8$ or increasing $\overleftarrow{n}_{22}$ beyond threshold $\overleftarrow{n}_{22}^*=3$  allows simultaneous improvement of both individual rates. Nonetheless, maximum improvement on $R_i$ is achieved by increasing $\overleftarrow{n}_{ii}$. $(b)$ Increasing either $\overleftarrow{n}_{11}$  or $\overleftarrow{n}_{22}$ beyond thresholds $\overleftarrow{n}_{11}^{*}$ and $\overleftarrow{n}_{22}^{*}$, allows maximum improvement of the sum rate (see Fig.~\ref{FigExample2a}).\\

\subsubsection{Example 3: At least one channel-output feedback link does not have any effect over the capacity region}\label{SecExample3}
\begin{figure}[t!]
\centerline{\epsfig{figure=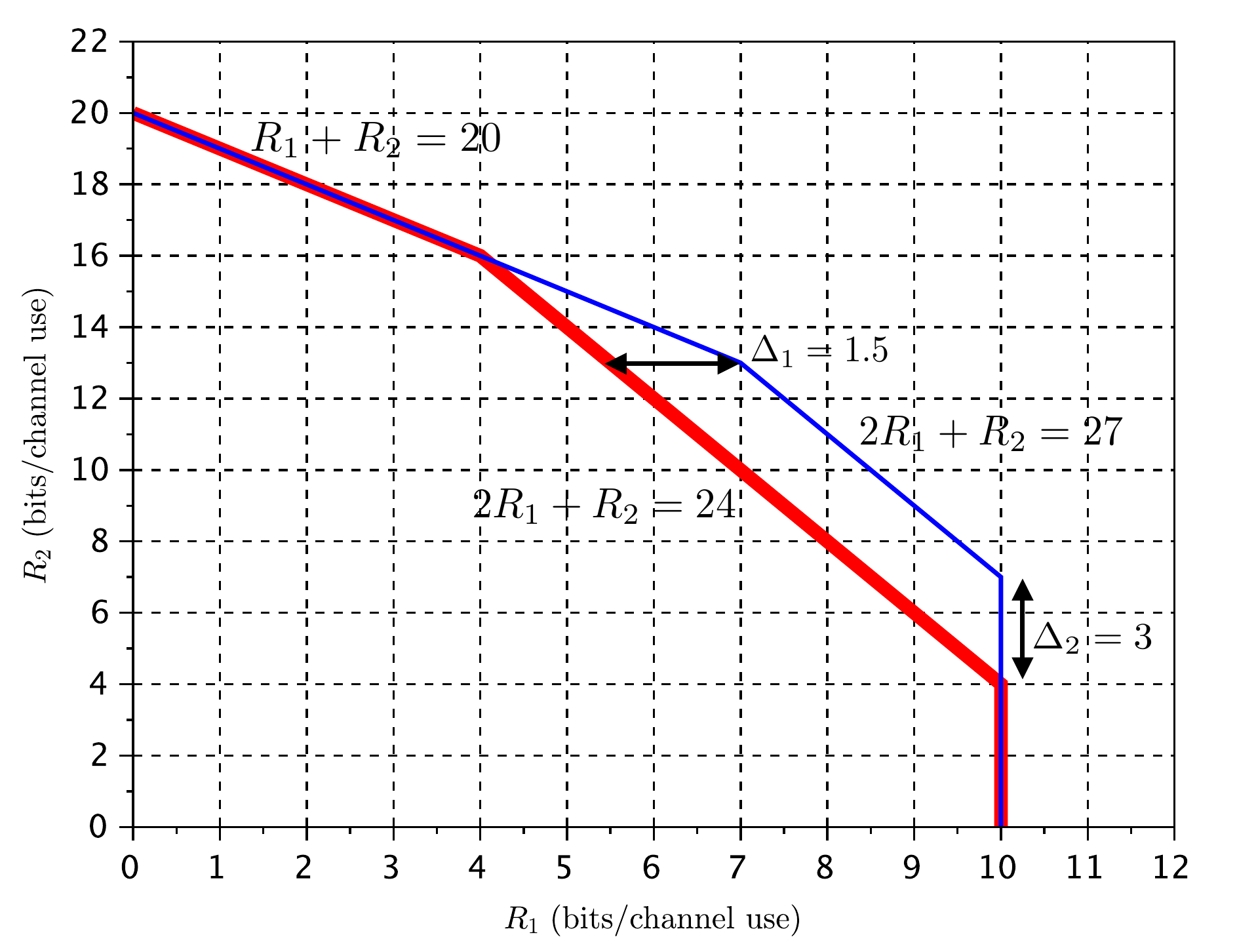,width=0.33\textwidth}}
\caption{Capacity region  $\Cldicnfb(10,20,6,12,0,0)$ without feedback (thick red line) and $\Cldicnfb(10,20,6,12,10,11)$ with noisy channel-output feedback (thin blue line) of the example in Sec.~\ref{SecExample3}.  Note that $\Delta_{1}(10,20,6,12,10,11) = 1.5$ bits/ch.use, $\Delta_{2}(10,20,6,12,10,11) = 2$ bits/ch.use and $\Sigma(10,20,6,12,10,11) = 0$ bits/ch.use.}
\label{FigExample3capb} 
\end{figure}
\begin{figure}[h]
\centerline{\epsfig{figure=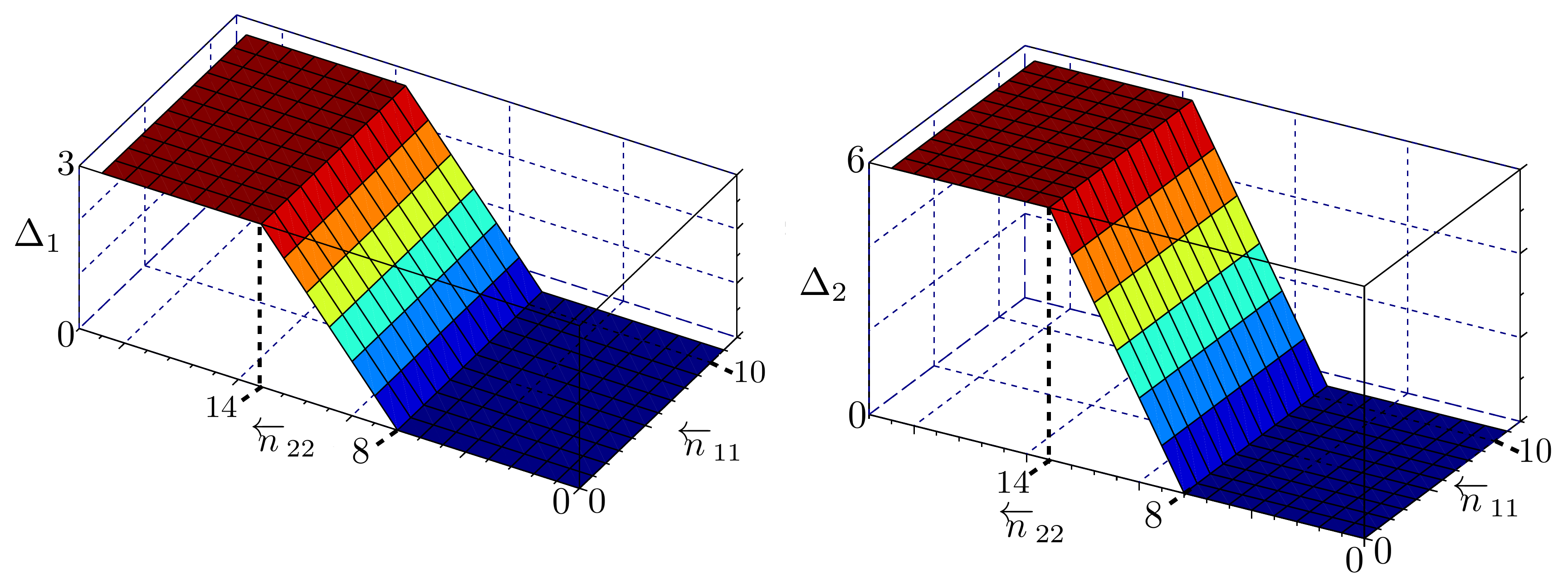,width=.43\textwidth}}
\caption{Maximum improvement of one individual rate of the example in Sec.~\ref{SecExample3}
%$\Delta_{i}(\overrightarrow{n}_{11}, \overrightarrow{n}_{22}, n_{12}, n_{21}, \overleftarrow{n}_{11} ,\overleftarrow{n}_{22})$, with $i \in \lbrace 1,2\rbrace$, as a function of $\overleftarrow{n}_{11}$ and $\overleftarrow{n}_{22}$ and fixed parameters $\overrightarrow{n}_{11} = $, $\overrightarrow{n}_{22} = $, $n_{12} = $, $n_{21} = $.
}
\label{FigExample3}
\end{figure}
Consider the case in which transmitter-receiver pairs $1$ and $2$ are in the weak interference regime, with ${\overrightarrow{n}_{11} = 10}$, ${\overrightarrow{n}_{22} = 20}$, ${n_{12} = 6}$, ${n_{21} = 12}$. 
In Fig.~\ref{FigExample3capb} the capacity region is plotted without channel-output feedback and with noisy channel-output feedback ($\overleftarrow{n}_{11}=10, \overleftarrow{n}_{22}=11$).
In Fig.~\ref{FigExample3}, $\Delta_{i}(10, 20, 6, 12, \overleftarrow{n}_{11} ,\overleftarrow{n}_{22})$ is plotted for both $i = 1$ and $i=2$ as a function of $\overleftarrow{n}_{11}$ and  $\overleftarrow{n}_{22}$. 
Therein, it is shown that:
$(a)$ Increasing parameter $\overleftarrow{n}_{11}$ does not enlarge the capacity region, independently of the value of $\overleftarrow{n}_{22}$.
$(b)$ Increasing parameter $\overleftarrow{n}_{22}$ beyond threshold $\overleftarrow{n}_{22}^*=8$ allows simultaneous improvement of both individual rates.
$(c)$ Finally, none of the parameters $\overleftarrow{n}_{11}$ or $\overleftarrow{n}_{22}$ increases the sum-rate in this case.

\subsubsection{Example 4: The channel-output feedback of link $i$ exclusively  improves $R_j$}\label{SecExample4}
\begin{figure}[t!]
\centerline{\epsfig{figure=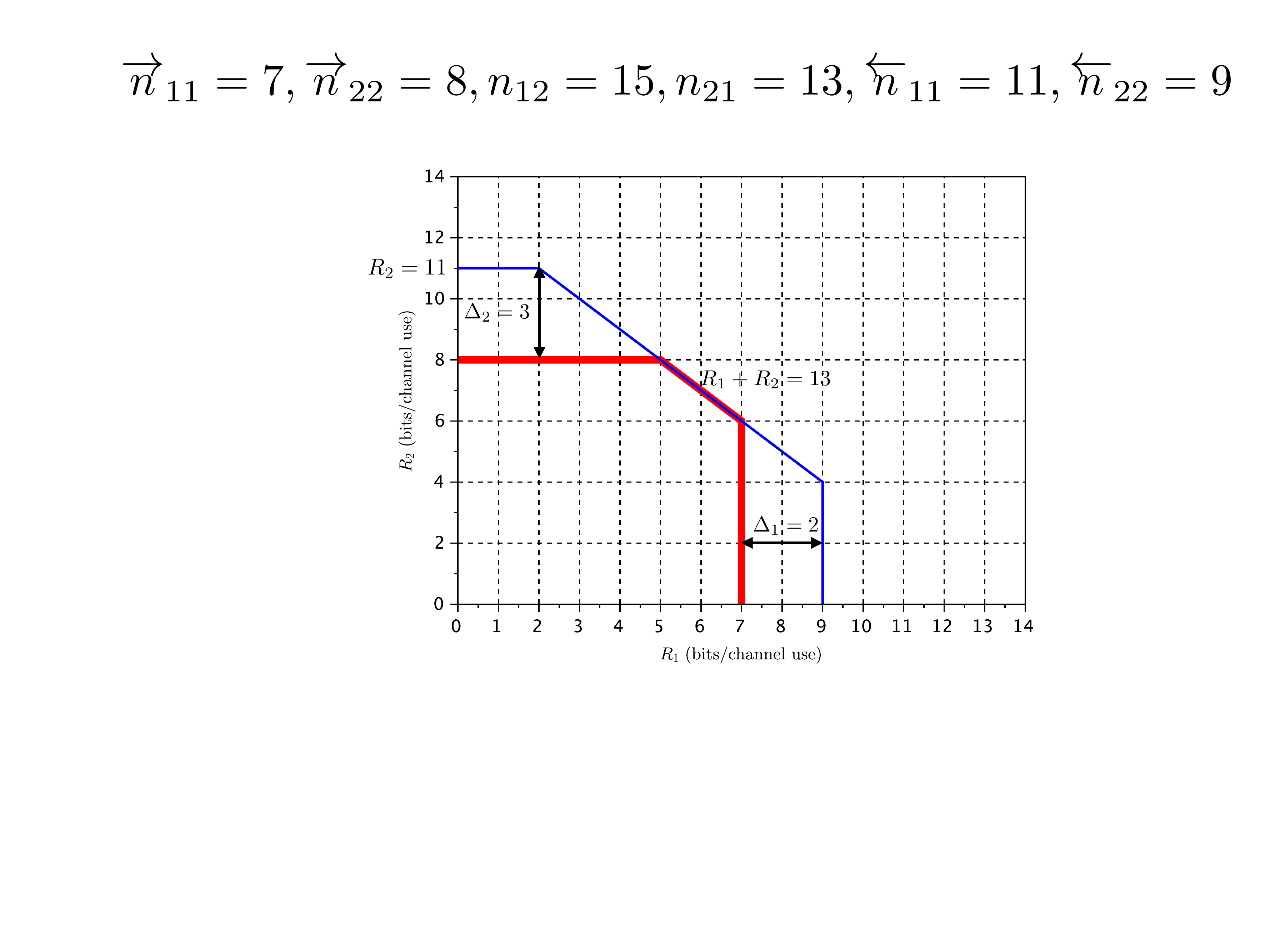,width=0.33\textwidth}}
\caption{Capacity region $\Cldicnfb(7,8,15,13,0,0)$ without feedback  (thick red line) and $\Cldicnfb(7,8,15,13,11,9)$ with noisy channel-output feedback  (thin blue line) of the example in Sec.~\ref{SecExample4}.  Note that $\Delta_{1}(7,8,15,13,11,9) = 2$ bits/ch.use, $\Delta_{2}(7,8,15,13,11,9) = 3$ bits/ch.use and $\Sigma(7,8,15,13,11,9) = 0$ bits/ch.use.}
\label{FigExample4capb}
\end{figure}
\begin{figure}[h]
\centerline{\epsfig{figure=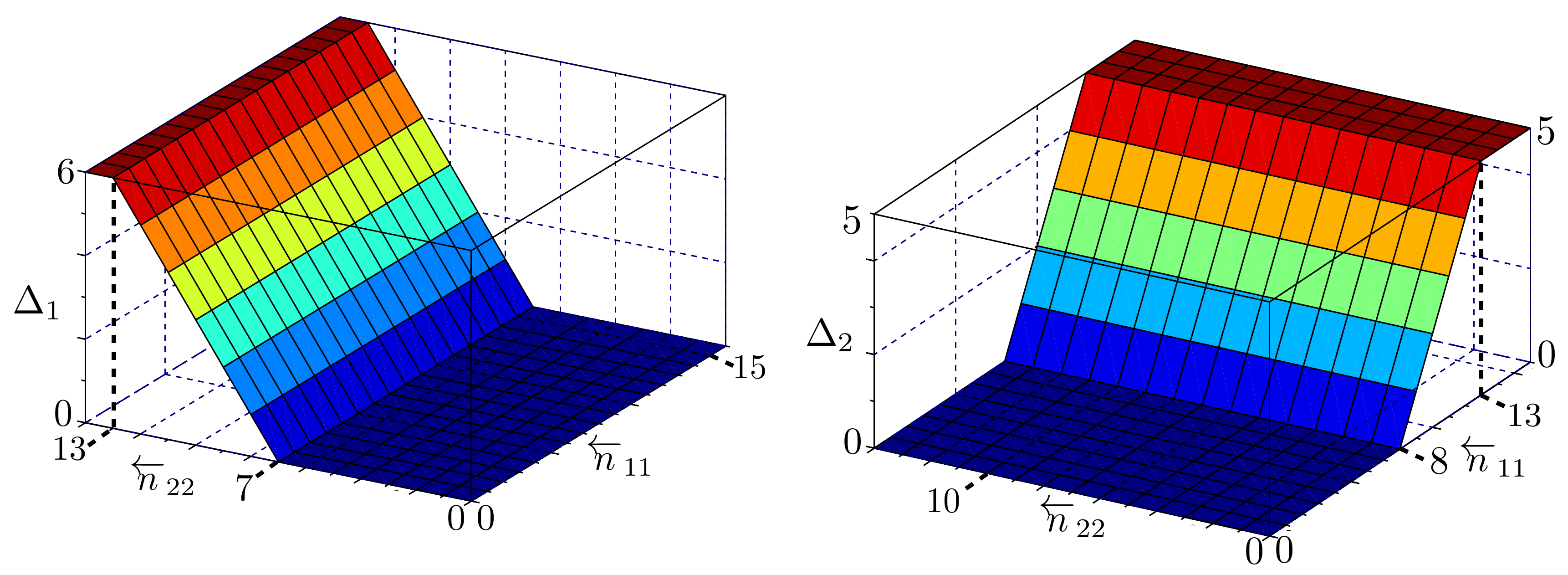,width=.43\textwidth}}
\caption{Maximum improvement of one individual rate of the example in Sec.~\ref{SecExample4}
%$\Delta_{i}(\overrightarrow{n}_{11}, \overrightarrow{n}_{22}, n_{12}, n_{21}, \overleftarrow{n}_{11} ,\overleftarrow{n}_{22})$, with $i \in \lbrace 1,2\rbrace$, as a function of $\overleftarrow{n}_{11}$ and $\overleftarrow{n}_{22}$ and fixed parameters $\overrightarrow{n}_{11} = $, $\overrightarrow{n}_{22} = $, $n_{12} = $, $n_{21} = $.
}
\label{FigExample4a}
\end{figure}
Consider the case in which transmitter-receiver pairs $1$ and $2$ are in the very strong and strong interference regimes, with $\overrightarrow{n}_{11} = 7$, $\overrightarrow{n}_{22} = 8$, $n_{12} = 15$, $n_{21} = 13$. 
In Fig.~\ref{FigExample4capb} the capacity region is plotted without channel-output feedback and with noisy channel-output feedback ($\overleftarrow{n}_{11}=11, \overleftarrow{n}_{22}=9$).
In Fig.~\ref{FigExample4a}, $\Delta_{i}(7, 8, 15, 13, \overleftarrow{n}_{11} ,\overleftarrow{n}_{22})$ is plotted for both $i = 1$ and $i=2$ as a function of $\overleftarrow{n}_{11}$ and  $\overleftarrow{n}_{22}$. 
Therein, it is shown that:
$(a)$ Increasing parameter $\overleftarrow{n}_{11}$ beyond threshold $\overleftarrow{n}_{11}^*=8$ exclusively improves $R_2$.
$(b)$ Increasing parameter $\overleftarrow{n}_{22}$ beyond threshold $\overleftarrow{n}_{22}^*=7$ exclusively improves $R_1$.
$(c)$ None of the parameters $\overleftarrow{n}_{11}$  or $\overleftarrow{n}_{22}$  has an impact over the sum rate in this case.
Note that these observations are in line with the interpretation of channel-output feedback as an altruistic technique, as in \cite{Perlaza-TIT-2013, Perlaza-ISIT-2014a}.  This is basically because the link implementing channel-output feedback provides an alternative path to the information sent by the other link, as first suggested in \cite{Suh-TIT-2011}.\\
\subsubsection{Example 5: None of the channel-output feedback links has any effect over the capacity region}
Consider the case in which transmitter-receiver pairs $1$ and $2$ are in the very weak and strong interference regimes, with $\overrightarrow{n}_{11} = 10$, $\overrightarrow{n}_{22} = 9$, $n_{12} = 2$, $n_{21} = 15$. 
Note that the capacity region of the LD-IC with and without channel-output feedback are identical, i.e., neither $\overleftarrow{n}_{11}$ nor $\overleftarrow{n}_{22}$ enlarges the capacity region.
In this specific example, it is not possible to take advantage of the feedback links for using interference as side information or to generate an alternative path.

\section{Conclusions}

In this paper, the noisy channel-output feedback capacity of the linear deterministic interference channel has been fully characterized by generalizing existing results. 
Based on specific asymmetric examples, it is highlighted that even in the presence of noise, the benefits of channel-output feedback can be significantly relevant in terms of achievable individual rate and sum-rate improvements with respect to the case without feedback. Nonetheless, there also exist scenarios in which these benefits are totally inexistent.
\balance

\balance
\bibliographystyle{IEEEtran}
\bibliography{IT-GT}

\begin{thebibliography}{10}
\providecommand{\url}[1]{#1}
\csname url@samestyle\endcsname
\providecommand{\newblock}{\relax}
\providecommand{\bibinfo}[2]{#2}
\providecommand{\BIBentrySTDinterwordspacing}{\spaceskip=0pt\relax}
\providecommand{\BIBentryALTinterwordstretchfactor}{4}
\providecommand{\BIBentryALTinterwordspacing}{\spaceskip=\fontdimen2\font plus
\BIBentryALTinterwordstretchfactor\fontdimen3\font minus
  \fontdimen4\font\relax}
\providecommand{\BIBforeignlanguage}[2]{{%
\expandafter\ifx\csname l@#1\endcsname\relax
\typeout{** WARNING: IEEEtran.bst: No hyphenation pattern has been}%
\typeout{** loaded for the language `#1'. Using the pattern for}%
\typeout{** the default language instead.}%
\else
\language=\csname l@#1\endcsname
\fi
#2}}
\providecommand{\BIBdecl}{\relax}
\BIBdecl

\bibitem{Tuninetti-ISIT-2007}
D.~Tuninetti, ``On interference channel with generalized feedback ({IFC-GF}),''
  in \emph{Proc. of International Symposium on Information Theory (ISIT)},
  Nice, France, Jun. 2007, pp. 2661--2665.

\bibitem{Tuninetti-ITA-2010}
------, ``An outer bound region for interference channels with generalized
  feedback,'' in \emph{IEEE Information Theory and Applications Workshop
  (ITA)}, Feb. 2010, pp. 1--5.

\bibitem{Suh-TIT-2011}
C.~Suh and D.~N.~C. Tse, ``Feedback capacity of the {G}aussian interference
  channel to within 2 bits,'' \emph{IEEE Transactions on Information Theory},
  vol.~57, no.~5, pp. 2667--2685, May. 2011.

\bibitem{Vahid-TIT-2012}
A.~Vahid, C.~Suh, and A.~S. Avestimehr, ``Interference channels with
  rate-limited feedback,'' \emph{IEEE Transactions on Information Theory},
  vol.~58, no.~5, pp. 2788--2812, May. 2012.

\bibitem{Yang-Tuninetti-TIT-2011}
S.~Yang and D.~Tuninetti, ``Interference channel with generalized feedback
  (a.k.a. with source cooperation): Part {I}: Achievable region,'' \emph{IEEE
  Transactions on Information Theory}, vol.~5, no.~57, pp. 2686--2710, May.
  2011.

\bibitem{Mohaher-TIT-2013}
S.~Mohajer, R.~Tandon, and H.~V. Poor, ``On the feedback capacity of the fully
  connected-user interference channel,'' \emph{IEEE Transactions on Information
  Theory}, vol.~59, no.~5, pp. 2863--2881, May. 2013.

\bibitem{Prabhakaran-TIT-2011}
V.~M. Pabhakaran and P.~Viswanath, ``Interference channel with source
  cooperation,'' \emph{IEEE Transactions on Information Theory}, vol.~57,
  no.~1, pp. 156--186, Jan. 2011.

\bibitem{Tuninetti-ITW-2012}
D.~Tuninetti, ``An outer bound for the memoryless two-user interference channel
  with general cooperation,'' in \emph{IEEE Information Theory Workshop (ITW)},
  Sep. 2012, pp. 217--221.

\bibitem{Sahai-TIT-2013}
A.~Sahai, V.~Aggarwal, M.~Yuksel, and A.~Sabharwal, ``Capacity of all nine
  models of channel output feedback for the two-user interference channel,''
  \emph{IEEE Transactions on Information Theory}, vol.~59, no.~11, pp.
  6957--6979, 2013.

\bibitem{Perlaza-TIT-2013}
S.~M. Perlaza, R.~Tandon, H.~V. Poor, and Z.~Han, ``Perfect output feedback in
  the two-user decentralized interference channel,'' \emph{arXiv:1306.2878},
  2013.

\bibitem{SyQuoc-TIT-2015}
S.-Q. Le, R.~Tandon, M.~Motani, and H.~V. Poor, ``Approximate capacity region
  for the symmetric {G}aussian interference channel with noisy feedback,''
  \emph{IEEE Transactions on Information Theory}, vol.~61, no.~7, pp.
  3737--3762, Jul. 2015.

\bibitem{Quintero-INRIA-TechRep-2015}
V.~Quintero, S.~M. Perlaza, and J.-M. Gorce, ``Noisy channel-output feedback
  capacity of the linear deterministic interference channel,'' INRIA Grenoble -
  Rh{\^o}ne-Alpes, Tech. Rep. 456, Jan. 2015.

\bibitem{Berry-TIT-2011}
R.~A. Berry and D.~N.~C. Tse, ``Shannon meets {Nash} on the interference
  channel,'' \emph{IEEE Transactions on Information Theory}, vol.~57, no.~5,
  pp. 2821--2836, May. 2011.

\bibitem{Perlaza-ISIT-2014a}
S.~M. Perlaza, R.~Tandon, and H.~V. Poor, ``Symmetric decentralized
  interference channels with noisy feedback,'' in \emph{Proc. IEEE Intl.
  Symposium on Information Theory (ISIT)}, Honolulu, HI, USA, Jun. 2014.

\bibitem{Bresler-ETT-2008}
G.~Bresler and D.~N.~C. Tse, ``The two user {G}aussian interference channel: A
  deterministic view,'' \emph{European Transactions on Telecommunications},
  vol.~19, no.~4, pp. 333--354, Apr. 2008.

\end{thebibliography}
\balance

\end{document}